\documentclass{article}
\usepackage{amsmath}                             
\usepackage{amsfonts}
\input psfig.sty

\def\Bi{/home/Brockett/navin/Bibliography}
\newtheorem{theorem}{Theorem}

\newtheorem{corollary}{Corollary}

\newtheorem{lemma}{Lemma}
\newtheorem{example}{Example}
\newtheorem{definition}{Definition}

\newtheorem{remark}{Remark}
\newtheorem{notation}{Notation}

\def\Iden{\mbox{$\bf 1\ $}}
\def\n{\noindent}

\def\h {\mathfrak{h}}
\def\m {\mathfrak{m}}
\def\g {\mathfrak{g}}
\def\k {\mathfrak{k}}
\def\s {\mathfrak{s}}

\def\a {\mathfrak{a}}
\def\b {\mathfrak{b}}
\def\f {\mathfrak{f}}

\def\su {\mathfrak{su}}
\def\u {\mathfrak{u}}

\def\Re {\mathbb{R}}

\def\hi {\mathbb{H}}

\begin{document}

\author{Navin Khaneja,\thanks{Department of Mathematics, Dartmouth College, Hanover, NH 03755, email: Navin.Khaneja@dartmouth.edu This work was supported by Burke Fellowship from Dartmouth College}\ \ \ Steffen J. Glaser\thanks{Institute
of Organic Chemistry and Biochemistry II, Technical University Munich,
85747 Garching, Germany. This work was funded by the Fonds der Chemischen Industrie and the Deutsche Forschungsgemeinschaft under grant Gl 203/1-6.}}

\title{{\bf Cartan Decomposition of SU$(2^n)$, Constructive Controllability of Spin Systems and Universal Quantum Computing}}

\maketitle

\begin{center}
{\bf Abstract}
\end{center}

\n In this paper we provide an explicit construction of any arbitrary unitary transformation on $n$ qubits from one qubit and a single two qubit quantum gate. Building on the previous work demonstrating the universality of two qubit quantum gates, we present here an explicit implementation. The construction is based on the Cartan decomposition of the semi-simple Lie group $SU$($2^n)$ and uses the geometric structure of the Riemannian Symmetric Space $\frac{SU(2^n)}{SU(2^{n-1}) \otimes SU(2^{n-1}) \otimes U(1)}$. The decomposition highlights the geometric aspects of the problem of building an arbitrary unitary transformations out of quantum gates and makes explicit the choice of pulse sequences for the implementation of arbitrary unitary transformation on $n$ coupled $\frac{1}{2}$ spins in $NMR$ quantum computing. Finally we make observations on the optimality of the design procedure. 

\begin{center}
\section{Introduction}\end{center}
Recent interests in quantum computing have been fostered by the prospects of 
building a computational theory which is different and more powerful than the classical computational theory. The vision of solving a class of problems using quantum computing, which have been known to be computationally hard in the classical setting, has been advanced by the recent work of \cite{shor}, \cite{grover}. Any quantum computation involves evolution of the initial state under a series of unitary transformations. Absolutely essential to the design of an universal quantum computer is the ability to synthesize any arbitrary unitary transformations from simple quantum gates (lower dimensional unitary transformations). The search for such universal components was initiated by Deutsch's original discovery \cite{deutsch} of a three-bit universal quantum logic gate. Following which DiVincenzo \cite{Divincenzo} showed that two-bit universal quantum gates also exist. This result was then generalized by the work of Barenco et al. \cite{barenco.et.al}, Lloyd \cite{lloyd} and Deutsch et al. \cite{deutsch} showing that almost any two bit gate is also universal. The central idea behind demonstrating the universality of these gates rests on the fact that the generators of the unitary transformations produced by these gates generate the Lie algebra of the unitary group and therefore by suitable composition of these gates, we can produce any arbitrary unitary transformation. This is basically the notion of controllability in mathematical control theory-- that is, whether available Hamiltonians can prepare an arbitrary state of the quantum system \cite{rama, tarn}. The conditions of controllability of a quantum system were rediscovered as the conditions for the universality of a quantum computer. 

\n Problems of similar nature arise in coherent spectroscopy. Many areas of spectroscopic fields, such as nuclear magnetic resonance (NMR), electron magnetic resonance and  optical
spectroscopy rely on a limited set of control variables in
order to create desired unitary transformations \cite{Ernst, Science, optics}.
In NMR, unitary transformations are used to
manipulate an ensemble of nuclear spins, e.g. to transfer
coherence between coupled spins in multidimensional
NMR-experiments \cite{Ernst} or to implement quantum-logic
gates in NMR quantum computers \cite{QC}. However, the
design of a sequence of radio-frequency pulses that
generate a desired unitary operator is not trivial \cite{design}. So far, no general design approach is known for the implementation of a
desired unitary transformation \cite{Science}. During the last decade the questions of controllability of quantum systems have generated considerable interest \cite{judson,herbruggen}. In particular, coherence or polarization transfer in pulsed coherent spectroscopy has received lot of attention \cite{Science, design}. Algorithms for determining bounds quantifying the maximum possible efficiency of transfer between non-Hermitian operators have been determined \cite{Science}. There is utmost need for design strategies for pulse sequences that can achieve these bounds. From a control theory perspective, all these are constructive controllability problems \cite{khaneja}. 

\n In this paper, we present a constructive solution to the problem of producing an arbitrary $2^n$ dimensional unitary transformation acting on $n$ qubits from unitary transformations acting on individual and pair of qubits (elements of $SU(2)$ and $SU(4)$ respectively). The design strategy presented here is a generalization of well known Euler angle decomposition for rotations. Recall if 

\begin{eqnarray*}
I_x &=& \frac{1}{2}\left(  
\begin{array}{cc}
 0 & 1\\
1 & 0 
\end{array}
\right)
\\  
I_y
 &=& \frac{1}{2}\left(  
\begin{array}{cc}
 0 & -i\\
i & 0 
\end{array}
\right)\\
I_z &=& \frac{1}{2}\left(  
\begin{array}{cc}
 1 & 0\\
0 & -1 
\end{array}
\right)
\end{eqnarray*}represent the standard Pauli spin matrices, then any $V \in SU(2)$ has a decomposition $$ V = \exp(-i\alpha I_x) \exp(-i\beta I_z) \exp(-i \gamma I_x).$$where $\alpha, \beta, \gamma \in \Re$. Similarly any $W \in SU(2^n)$, has a decomposition $$W = K_1AK_2.$$ Where $K_1, K_2 \in SU(2^{n-1}) \otimes SU(2^{n-1}) \otimes U(1)$ and $A \in \exp(\h)$, where $\h$ is a Cartan subalgebra of the Riemannian symmetric space $$\frac{SU(2^n)}{SU(2^{n-1}) \otimes SU(2^{n-1}) \otimes U(1)}.$$

\n We will elaborate on all these notions. The point to be emphasized here is that we then obtain a recursive formula for the decomposition. Given the decomposition $W = K_1AK_2$, we can further decompose $K_1$ and $K_2$ in terms of the elements of $SU(2^{n-2}) \otimes SU(2^{n-2}) \otimes U(1)$ and so on. Finally we will be left with only one and two qubit operations. We will see that this recursive formulation highlights the geometric structure of the problem. 

We are also interested in the implementation of these unitary transformations in a network of coupled spins in the context of NMR. The emphasis being NMR quantum computers and coherence transfer experiments in multidimensional spectroscopy. We emphasize again that from a control theory viewpoint, these are constructive controllability problems. The dynamical system to be controlled is defined through the time-dependent Schr{\"o}dinger equation $$\dot{U(t)} = -iH(t)U(t),\ \ U(0)=I,$$ where $H(t)$ and $U(t)$ are the Hamiltonian and the unitary displacement operators, respectively. In this paper, we will only be concerned with finite-dimensional quantum systems. In this case, we can choose a basis and think of $H(t)$ as a Hermitian matrix. We can split the Hamiltonian $$H = H_d + \sum_{i=1}^m v_i(t)H_i,$$ where $H_d$ is the part of Hamiltonian that is internal to the system and we call it the {\it drift} or {\it free evolution Hamiltonian} and $\sum_{i=1}^m v_i(t)H_i$ is the part of Hamiltonian that can be externally changed. It is called the {\it control} or {\it rf Hamiltonian}. The equation for $U(t)$ dictates the evolution of the density matrix according to $$\rho(t) = U(t)\rho(0)U^{\dagger}(t).$$ where \begin{equation}\label{eq:unitary}\dot{U} = -i(H_d + \sum_{i=1}^mv_iH_i)\ U; U(0)=I\ .\end{equation} In a network of coupled spin $\frac{1}{2}$ nuclei, the control terms $H_i$ correspond to the Hamiltonian that effect each spin individually (we assume that resonance frequencies of the spins are well separated so that selective excitation is possible). The group $K$ generated by the Lie algebra $\{H_i\}_{LA}$ corresponds to $n$ direct copies of $SU(2)$, which we denote by $$ \left[ SU(2) \right]^{\otimes n}.$$ The drift term $H_d$ is the part of the Hamiltonian that corresponds to couplings among the spins. It is known that if the Lie algebra $\{H_d, H_i \}_{LA}$ equals $\su(2^n)$, the algebra of $2^n \times 2^n$, traceless skew Hermitian matrices, then we can steer the system ($\ref{eq:unitary}$), from identity to any point $U_F \in SU(2^n)$. The problem we will address in this paper is explicit construction of a pulse sequence or in other words control laws $v_i$, that steer the system (\ref{eq:unitary}) from $U(0)= I$ to some $U_F$, in finite time. 

\n We begin by reviewing some important facts about Lie groups and Lie algebras, which will be used in the remaining part of the paper. The exposition is very brief and follows \cite{gilmore}. The reader is advised to refer to \cite{nomizu, Helg, wolf} for more details and a rigorous treatment on the subject. 

\begin{center}
\item \section{Cartan Decomposition and Riemannian Symmetric Spaces}\label{sec:prem} \end{center}

\n We will assume that the reader is familiar with the basic facts about Lie groups and homogeneous spaces \cite{nomizu}. 

\begin{notation}{\rm Throughout the paper, $G$ will denote a compact semi-simple Lie group and $e$ its identity element (we use I to denote the identity matrix when working with the matrix representation of the group). As is well known there is a naturally defined bi-invariant metric on $G$, given by the Killing form. We denote this bi-invariant metric by $<,>_{G}$. We will use $K$ to denote a compact closed subgroup of $G$. Let $L(G)$ be the Lie algebra of right invariant vector fields on $G$ and similarly $L(K)$ the Lie algebra of right invariant vector fields on $K$. There is a one to one correspondence between these vector fields and the tangent spaces $T_e(G)$ and $T_e(K)$, which we denote by $\g$ and $\k$ respectively. There is a direct sum decomposition $\g = \m \oplus \k$ such that $\m = \k^{\bot}$ with respect to the metric.}
\end{notation} 

\n To fix ideas, let $G = SU(n)$ and $\g = \su (n)$ be its associated Lie algebra of $n \times n$ traceless skew-Hermitian matrices . Then $<A,B>_{G} = trace(A^{\dagger}B),\ A,B \in \su (n)$ (which is proportional to the Killing metric) represents a bi-invariant metric on $SU(n)$.

\begin{definition} {\bf (Cartan decomposition of $\g$)} {\rm Let $\g$ be a real semi-simple Lie algebra and let the decomposition $\g = \m \oplus \k$, $\m = \k^{\bot}$ satisfy the commutation relations 
\begin{eqnarray}
\left[ \k,\k \right] &\subset& \k \\
\left[ \m,\k \right] &=& \m \\
\left[ \m,\m \right] &\subset& \k. 
\end{eqnarray} We will refer to this decomposition as a Cartan decomposition of $\g$. The pair $(\g, \k)$ will be called an orthogonal symmetric Lie algebra pair \cite{gilmore,Helg} }.  
\end{definition}

\begin{definition}\label{def:adjoint.orbit}{\bf(Adjoint orbit)}{\rm The Lie group $G$ acts on its Lie algebra $\g$ by conjugation $Ad_G : \g \rightarrow \g$ (called the adjoint action). This is defined as follows. Given $U \in G, \ X \in \g$, then $$Ad_U(X) = \frac{d\ U \exp(tX)U^{-1}}{dt}|_{t=0}.$$ We use the notation $$Ad_K(X) = \bigcup_{k \in K}Ad_k(X).$$ $Ad_K(X)$ is called the adjoint orbit of $X$.}
\end{definition}

\n For example if $U \in SU(n)$, $A \in \su (n)$, its associated Lie algebra of $n \times n$ traceless skew-Hermitian matrices . Then $Ad_{U}(A) = UAU^{T}$, such that $$ \exp(Ad_{U}(A)) = U \exp(A) U^{T} = Ad_{U}(\exp(A)). $$

\begin{definition} {\bf(Cartan subalgebra)} {\rm Consider the semi-simple Lie algebra $\g$ and its Cartan decomposition $\g = \m \oplus \k$. If $\h$ is a subalgebra of $\g$ contained in $\m$, then $\h$ is Abelian because $[\m, \m] \in \k$. A maximal Abelian subalgebra contained in $\m$ is called a Cartan subalgebra of the pair $(\g, \k)$.} 
\end{definition}

\n It is well known \cite{Helg, wolf} that:

\begin{theorem}\label{th:rank}{\rm If $\h$ and $\h'$ are two maximal Abelian subalgebras contained in $\m$, then 
\begin{enumerate}

\item There is an element $k \in K$ such that $Ad_k(\h) = \h'$.

\item $\m = \bigcup_{k \in K} Ad_k(\h)$. 
\end{enumerate}}
\end{theorem}

\begin{remark} {\rm It is also well known that the homogeneous coset space $G/K = \{KU : U \in G \}$ admits the structure of a differentiable manifold \cite{nomizu}. Let $\pi: G \rightarrow G/K$ denote the natural projection map. Define $o \in G/K$ by $o = \pi(e)$. Given the decomposition $\g = \m \oplus \k$, the tangent space plane $T_{o}(G/K)$ can be then identified with the vector subspace $\m$. If $\g = \m \oplus \k$ is a Cartan decomposition then the homogeneous space $G/K = \exp(\m)$, and is called a {\it globally Riemannian symmetric space} \cite{wolf}. From above stated theorem \ref{th:rank}, the maximal Abelian subalgebras of $\m$ are all $Ad_K$ conjugate and in particular they have the same dimension. The dimension is called the {\it rank} of the globally Riemannian symmetric space $G/K$.}
\end{remark}

\begin{theorem}\label{th:decomp}{\rm Given the semi-simple Lie algebra $\g$ and its cartan decomposition $\g = \m \oplus \k$, let $\h$ be a Cartan subalgebra of the pair $(\g,\k)$ and define $A = \exp (\h) \subset G$. Then $G = KAK$.}
\end{theorem}
\n {\bf Proof:} $G = KP$, where $P = \exp(\m)= \exp(\bigcup_{k \in K} Ad_k(\h)) = \bigcup_{k \in K}Ad_k(\exp (\h)) = \bigcup_{k \in K}Ad_k(A) \subset KAK$. Now $G = KKAK = KAK$. \hfill{\bf{Q.E.D}}.  
 
\n \begin{definition} {\bf (Cartan decomposition of $G$)} {\rm The decomposition $G = KAK$ of the semi-simple Lie group $G = KAK$, will be our most important tool in this paper. We will call this decomposition the {\it Cartan decomposition} of the Lie group $G$ \cite{Helg}}
\end{definition}

\n \begin{definition} {\bf (Weyl orbit)} {\rm Given the Cartan decomposition $\g = \m \oplus \k$, let $\h$ be a Cartan subalgebra of $(\g, \k)$ containing $X$. We use the notation $W(X)= \h \bigcap Ad_K(X)$ to denote the maximal commuting set contained in the adjoint orbit of $X$. The set $W(X)$ is called the Weyl orbit of $X$.}
\end{definition}

\begin{center} \section{Product Operator Basis} \label{sec:product} \end{center} 
The Lie Group $G$ which we will be most interested in is $SU(2^{n})$, the special unitary group describing the evolution of $n$ interacting spin $\frac{1}{2}$ particles ( Please note that we focus on $SU(2^{n})$ instead of $U(2^n)$ because a global phase is not of interest to us). The Lie algebra $\su (2^n)$ is a $4^n -1$ dimensional space of traceless $n \times n$ skew-Hermitian matrices. The orthonormal bases which we will use for this space is expressed as tensor products of Pauli spin matrices \cite{sorenson}(product operator bases). We choose to work in these bases because of their widespread use in the NMR literature and our desire to look at the implementations of NMR quantum computers. Recall the Pauli spin matrices $I_x$, $I_y$, $I_z$ defined by

\begin{eqnarray*}
\label{pauli}
I_x &=& \frac{1}{2}\left(  
\begin{array}{cc}
 0 & 1\\
1 & 0 
\end{array}
\right)
\\  
I_y &=& \frac{1}{2}\left(  
\begin{array}{cc}
 0 & -i\\
i & 0 
\end{array}
\right)\\
I_z &=& \frac{1}{2}\left(  
\begin{array}{cc}
 1 & 0\\
0 & -1 
\end{array}
\right)
\end{eqnarray*}are the generators
of the rotation in the two dimensional Hilbert space and  basis for the Lie algebra of
traceless skew-Hermitian matrices $\su (2)$. They obey the well known relations
\begin{equation}\label{eq:pauli.1}
 [I_x, \ I_y] = i I_z \; \; ; \; \; [I_y ,\ I_z] = i I_x \; \; ; \;
\;   [I_z, \ I_x] = i I_y \end{equation}
\begin{equation}\label{eq:pauli.2}
I_x^2 = I_y^2 = I_{z}^2 = \frac{1}{4}\Iden 
\end{equation}where $$ \Iden = \left(  
\begin{array}{cc}
 1 & 0\\
0 & 1 
\end{array}
\right)$$

\begin{notation} {\rm The orthogonal basis $\{iB_s\}$, for $\su (2^n)$ take the form 
\begin{equation}
B_s = 2^{q-1}\prod_{k=1}^{n}(I_{k\alpha})^{a_{ks}},
\end{equation}$\alpha = x, y, or \ z$ and  
\begin{equation}
\label{tensor}
 I_{k\alpha} = \Iden \otimes \cdots \otimes I_{\alpha} \otimes \Iden,
\end{equation}
where $I_{\alpha}$ the Pauli matrix appears in the above expression only at the $k^{th}$ 
position, and $\Iden $ the two dimensional identity matrix, appears everywhere 
except at the $k^{th}$ position. $a_{ks}$ is $1$ in $q$ of the indices and $0$ in the remaining. Note that $q \geq 1$ as $q = 0$ corresponds to the identity matrix and is not a part of the algebra. }
\end{notation}

\begin{example} {\rm As an example for $n=2$ the basis for $\su (4)$ takes the form 
\begin{eqnarray*}
q = 1 &\ & i\{I_{1x}, I_{1y}, I_{1z}, I_{2x}, I_{2y}, I_{2z} \}\\
q = 2 &\ & i\{I_{1x}I_{2x}, I_{1x}I_{2y}, I_{1x}I_{2z} \\ 
&\ & I_{1y}I_{2x},I_{1y}I_{2y}, I_{1y}I_{2z} \\
&\ & I_{1z}I_{2x},I_{1z}I_{2y}, I_{1z}I_{2z}.\} 
\end{eqnarray*}}
\end{example} 

\begin{remark} {\rm It is very important to note that the expression $I_{k\alpha}$ depends on the dimension $n$. To illustrate what this means, the expression for $I_{2z}$ for $n=2$ and $n=3$ is 
$\Iden \otimes I_z$, and $\Iden \otimes I_z \otimes \Iden$ respectively. Also observe that these operators are only normalized for $n=2$ as

\begin{equation}\label{eq:ortho}tr(B_r B_s) = \delta_{rs}2^{n-2}
\end{equation}}
\end{remark}

\n To fix ideas, lets compute one of these operators explicitly for $n=2$ 

\[I_{1z} = \frac{1}{2}\left[  
\begin{array}{cc}
1 & 0 \\
0 & -1
\end{array}
\right] \otimes \left[  
\begin{array}{cc}
1 & 0 \\
0 & 1
\end{array}
\right]   
\]  
which takes the form 
\[I_{1z} = \frac{1}{2}\left[  
\begin{array}{cccc}
1& 0 & 0 & 0\\
 0& 1 & 0 & 0\\
0& 0 & -1 & 0 \\
0 & 0 & 0 & -1
\end{array}
\right].   
\]

\begin{center}\section{The Two Qubit Example}\end{center}
Before we consider the most general case of $n$ qubits, let us make concrete the ideas developed in the previous section with the help of an example. 

\begin{example}\label{ex:2spin} {\rm Suppose we have two heteronuclear interacting qubits and the interaction between them produces a unitary transformation of the form $$ \exp (-i \alpha J I_{1z}I_{2z}),\ \ \alpha \in \Re. $$ This is a typical scenario of two nuclear spins coupled by a scalar J coupling. Furthermore assume we can individually excite each spin, i.e. perform one qubit operations. The goal now is to produce any arbitrary unitary transformation $U \in SU(4)$, from this specified coupling and one qubit operations. \n This structure appears often in the NMR situation. The unitary propogator $U$, describing the evolution of the system in a suitable rotating frame is described by \begin{equation} \label{eq:het.two.1} \dot{U} = -i (\ H_d + \sum_{i=1}^{4}u_i H_i \ )U , \ \ U(0)= I \end{equation} where \begin{eqnarray*}H_d &=& 2\pi J I_{1z} I_{2z} \\
H_1 &=& 2\pi I_{1x} \\
\ H_2 &=& 2\pi I_{1y} \\
\ H_3 &=& 2\pi I_{2x} \\
H_4 &=& 2\pi I_{2y},\end{eqnarray*}The symbol $J$ represents the strength of the scalar coupling between $I_1$ and $I_2$. Observe that the subgroup $K$ generated by $\{H_i \}_{i=1}^{4}$ is $SU(2)\otimes SU(2)$. Therefore the unitary transformations belonging to $SU(2)\otimes SU(2)$ can be produced very fast by hard pulses that excite each of the spins individually. 

The Lie algebra $\g = \su(4)$, has the direct sum decomposition $\g = \m \oplus \k$. Where

\begin{eqnarray*}
\k &=& {\mbox span} \ \ i\{ I_{1x}, I_{1y}, I_{1z}, I_{2x}, I_{2y}, I_{2z} \} \\
\m &=& {\mbox span} \ \ i\{ I_{1x}I_{2x}, I_{1x}I_{2y}, I_{1x}I_{2z} \\ 
&\ & \ \ \ \ \ I_{1y}I_{2x},I_{1y}I_{2y}, I_{1y}I_{2z} \\
&\ & \ \ \ \ \ I_{1z}I_{2x},I_{1z}I_{2y}, I_{1z}I_{2z} \} 
\end{eqnarray*}

\n Please note that span in above equations and in the rest of the paper denotes all linear combinations with real coefficients. Using the well known commutation relations $$[A \otimes B, C \otimes D] = [A, C]\otimes (B.D) + (C.A)\otimes [B,D],$$ and equations (\ref{eq:pauli.1}, \ref{eq:pauli.2}), it is easily verified $$[\k,\k] \subset \k ,\ [\m,\k] = \m ,\ [\m,\m] \subset \k .$$ Therefore the decomposition $\g = \m \oplus \k$ is a Cartan decomposition of $\su(4)$. As the subalgebra $\k = \su(2) \oplus \su(2)$, generates the group $K = SU(2) \otimes SU(2)$, the coset space $$\frac{SU(4)}{SU(2) \otimes SU(2)}$$ is a Riemannian symmetric space. Note that the Abelian subalgebra $\h$ generated by 
$$ i\{ I_{1x}I_{2x}, I_{1y}I_{2y}, I_{1z}I_{2z} \}$$ is contained in $\m$ and is maximal Abelian and hence a Cartan subalgebra of the Symmetric Space $\frac{SU(4)}{SU(2) \otimes SU(2)}$. Therefore using the corollary \ref{cor:decomp.0} any $U_F \in SU(4)$ can be decomposed as 

$$ U_F = K_1 \exp(-i (\alpha_1 I_{1x}I_{2x} + \alpha_2 I_{1y}I_{2y} + \alpha_3 I_{1z}I_{2z})) K_2 .$$ where $K_1, K_2 \in SU(2) \otimes SU(2)$. 

\n Now lets see how this decomposition makes obvious the choice of pulse sequences for producing this propogator. Note that for $K_y = \exp(-i\frac{\pi}{2} I_{1y})\exp(-i\frac{\pi}{2} {I_{2y}})$ , we have $$ K_y\exp(-i \alpha_1 I_{1z}I_{2z}) K_y^{-1} = \exp(-i \alpha_1 I_{1x}I_{2x})$$ and similarly for $K_x = \exp(-i\frac{\pi}{2} I_{1x})\exp(-i \frac{\pi}{2} I_{2x})$ we have $$ K_x^{-1} \exp(-i \alpha_2 I_{1z}I_{2z}) K_x = \exp(-i \alpha_2 I_{1y}I_{2y}). $$ This makes transparent, as to how one should generate the unitary transformation $U_F$ above using the unitary evolution in equation (\ref{eq:het.two.1}). This can be summarized by writing $U_F$ as \begin{equation} \label{2bit.decomp} U_F = K_1 \cdot K_y \cdot \exp(-i \alpha_1 I_{1z}I_{2z}) \cdot K_y^{-1}\cdot K_x^{-1} \cdot \exp (-i \alpha_2 I_{1z}I_{2z})\cdot K_x \cdot \exp (-i \alpha_3 I_{1z}I_{2z})) \cdot K_2 .\end{equation} The unitary propogators $K_x,K_y, K_1$ and $K_2$ can be produced by selective hard pulses, i.e. one qubit gates. 
}
\end{example}
\begin{remark} {\rm In \cite{time:khan}, it was shown that synthesizing $U_F$ using the decomposition given above, is indeed the fastest way to generate $U_F$. Time optimality is an important consideration in presence of decoherence. Observe that the Hamiltonians $I_{1x}I_{2x}$ and $I_{1y}I_{2y}$ are the elements of Weyl orbit of $I_{1z}I_{2z}$ under the adjoint action of $SU(2) \otimes SU(2)$. }
\end{remark}   

\n Now lets proceed to the general case of $n$ qubits.

\begin{center}
\item \section{ How does $[SU(2)]^{\otimes n}$ sit inside $SU(2^n)$} 
\end{center}

\n In this section we would like to use the machinery developed in the previous section to decompose the group $SU(2^n)$, into smaller unitary transformations, which can be produced. This will be achieved through successive Cartan decompositions of $SU(2^n)$ into smaller and smaller unitary transformations till we are only left with unitary operations corresponding to one and two bit operations.

\n {\bf Main Idea:} Suppose we are given $n$ qubits. In NMR quantum computing this will be a network of $n$ coupled spin $\frac{1}{2}$ particles. Let us label these qubits as $q_1, q_2 \dots q_n$. Let $\hi_1$ denote the two dimensional Hilbert space of a single qubit. Similarly
let $\hi_n$ denote the $2^n$ dimensional Hilbert space of $n$ qubits. These are related by $$ \hi_n = \hi_1 \otimes \hi_1 \otimes \dots \hi_1 .$$ 

\n To understand the decomposition of $SU(2^n)$, suppose we can generate an arbitrary unitary transformation on the Hilbert space $\hi_{n-1}$ of qubits $q_1, q_2 \dots q_{n-1}$ i.e. an arbitrary element of $SU(2^{n-1})$. Also assume, we can independently manipulate the $n^{th}$ qubit. Furthermore assume we have a two qubit gate that will act on the qubit $q_n$ and $q_{n-1}$. In the context of NMR, this corresponds to evolution under a coupling between the spin $q_n$ and $q_{n-1}$. To be more specific let this unitary evolution be of the form
$$\exp (-i\alpha J I_{(n-1)z}I_{nz}),$$ caused due to a scalar $J$ coupling.

We now explicitly build any $U \in SU(2^n)$ from elements of $SU(2^{n-1}) \otimes SU(2)$ and this interaction between $q_{n-1}$ and $q_n$. Notice this will essentially solve our main problem because to generate any element $V \in SU(2^{n-1})$ we will use this divide and conquer strategy again, building $V$ from $SU(2^{n-2}) \otimes SU(2)$ and interaction between $(n-1)^{th}$ and $(n-2)^{th}$ qubit and so on. Let us look at the geometry of the situation

\begin{notation}{\rm 
\n The Lie algebra $\su(2^{n})$ consists of the elements
$$ \su(2^{n}) = span \ \{ A\otimes I_{x}, B\otimes I_{y},  C\otimes I_{z},  D\otimes \Iden , iI_{nx},iI_{ny},iI_{nz} | A, B, C, D \in \su(2^{n-1}) \} $$   
The Lie algebra $\su(2^{n})$ has a direct sum decomposition into the following two vector spaces $$ \su_{\k}(2^{n}) = span\ \{ A\otimes I_{z},  B\otimes \Iden, iI_{nz} | A, B \in \su(2^{n-1}) \}, $$ $$ \su_{\m}(2^{n}) = span\ \{ A\otimes I_{x},  B\otimes I_{y}, iI_{nx}, iI_{ny} | A, B \in \su(2^{n-1}) \}. $$ We emphasize again that $span$ denotes the vector space generated by the elements of the set by taking linear combinations over the field of reals.} 
\end{notation} 

\begin{lemma} \label{lem:liesub}{ \rm The vector space $\su_{\k}(2^{n})$ is a Lie algebra of dimension $2 \times 4^{n-1} -1$ such that $$\exp(\su_{\k}(2^{n})) = SU(2^{n-1}) \otimes SU(2^{n-1}) \otimes U(1) $$} \end{lemma}

\n {\bf Proof:} The proof is a direct consequence of following commutation relations. Let $A, B, C \in \su(2^{n-1})$, if $A \neq B$ 
\begin{eqnarray*}
\left[ A\otimes I_{z} , B\otimes I_{z} \right] &=& C\otimes \Iden ;\\
\left[ A\otimes I_{z} , B\otimes \Iden \right] &=& C\otimes I_{z} ;\\
\left[ A\otimes \Iden , B\otimes \Iden \right] &=& C\otimes \Iden ;\\
\left[ A\otimes I_{z} , I_{nz}\right] &=& 0 ;\\
\left[ A\otimes \Iden , I_{nz} \right] &=& 0; \\
\left[ A\otimes I_{z} , A\otimes \Iden \right] &=& 0.
\end{eqnarray*}

\n Now observe that $I_{nz}$ and elements of the type 
$$ A\otimes (\frac{1}{2} \Iden + I_{z}) = A\otimes \left[  
\begin{array}{cc}
1& 0 \\
 0& 0\\
\end{array}
\right], $$ and
$$ B\otimes (\frac{1}{2} \Iden - I_{z}) = B\otimes \left[  
\begin{array}{cc}
0& 0 \\
 0& 1\\
\end{array}
\right].$$ commute.
Therefore the Lie Algebra  
$$ \su_{\k}(2^{n})= \su(2^{n-1}) \oplus \su(2^{n-1}) \oplus \u(1),$$
and the last part of the lemma follows. \hfill{\bf{Q.E.D}}.

\begin{theorem}{\rm The decomposition $\su(2^{n}) = \su_{\m}(2^{n}) \oplus \su_{\k}(2^{n})$ is a Cartan decomposition of the Lie algebra $\su(2^{n})$}
\end{theorem}

\n {\bf Proof:} From equation (\ref{eq:ortho}), it is clear that $\su_{\k}(2^{n}) \perp \su_{\m}(2^n)$. We have already shown in Lemma \ref{lem:liesub} that $\su_{\k}(2^{n})$ is a Lie subalgebra of $\su(2^{n})$. All that we need to show is that the commutation rules 
\begin{eqnarray*}
\left[\su_{\m}(2^{n}),\su_{\k}(2^{n}) \right] &=& \su_{\m}(2^{n}) \\
\left[\su_{\m}(2^{n}),\su_{\m}(2^{n}) \right] &\subset& \su_{\k}(2^{n}). 
\end{eqnarray*} are satisfied. The proof follows from the following commutation relations and the fact that $\su(n)$ is semi-simple (implies $[\su(n), \su(n)] = \su(n)$). Let $A, B, C \in \su(2^{n-1})$, $c \in \Re$ and $\alpha \in \{ x,y \}$. If $A \neq B$, then 
\begin{eqnarray*}
\left[ A\otimes I_{\alpha}, B\otimes I_{\alpha} \right] &=& C\otimes \Iden ;\\
\left[ A\otimes I_{x} , B\otimes I_{y} \right] &=& C\otimes I_{z} ; \\
\left[ A\otimes I_{x} , B\otimes I_{z} \right] &=& C\otimes I_{y} ; \\
\left[ A\otimes I_{y} , B\otimes I_{z} \right] &=& C\otimes I_{x} ; \\
\left[ A\otimes I_{\alpha} , B\otimes \Iden \right] &=& C\otimes I_{\alpha}. \\
\end{eqnarray*}
If $A = B$ then
\begin{eqnarray*}
\left[ A\otimes I_{x} , A\otimes I_{y} \right] &=& c\ iI_{nz} ; \\
\left[ A\otimes I_{x} , A\otimes I_{z} \right] &=& c\ iI_{ny} ; \\
\left[ A\otimes I_{y} , A\otimes I_{z} \right] &=& c\ iI_{nx} ; \\
\left[ A\otimes I_{\alpha} , A\otimes \Iden \right] &=& 0.
\end{eqnarray*}

 \begin{corollary}\label{cor:decomp.0}{\rm Any $U \in SU(2^n)$ has a decomposition $$U = K_1 A K_2,$$ where $ K_1, K_2 \in \exp(\su_{\k}(2^n)) \cong SU(2^{n-1}) \otimes SU(2^{n-1}) \otimes U(1)$ and $A = \exp (Y)$, for some $Y \in \h$, a maximal Cartan subalgebra of the pair $(\su(2^{n}), \su_{\k}(2^{n}))$.} 
\end{corollary}

\n {\bf Proof:} We have already shown that $\su(2^{n}) = \su_{\k}(2^{n}) \oplus \su_{\m}(2^{n})$ is a Cartan decomposition. Hence the proof follows from theorem \ref{th:decomp}. \hfill{\bf{Q.E.D}}.

\n The key observation is that $ K_1, K_2 \in \exp(\su_{\k}(2^n)) $ again has a decomposition. This is stated in the following theorem.

\n \begin{notation}{\rm Let us denote 
\begin{eqnarray*}
\su_{\k1}(2^{n})&=& span\ \{ A\otimes I_{z}| A \in \su(2^{n-1}) \} \\
\su_{\k0}(2^{n})&=& span\ \{ A\otimes \Iden| A \in \su(2^{n-1}) \} \\
\overline{\su_{\k}(2^{n})} &=& span\ \{ A\otimes \Iden,  A\otimes I_{z} | A \in \su(2^{n-1}) \}
\end{eqnarray*} Notice $\su_{\k}(2^{n})$ is just $\overline{\su_{\k}(2^{n})}\oplus \u(1)$ ($I_{nz}$ commutes with eveything in $\su_{\k}(2^{n})$ and generates $\u(1)$) and therefore $$\exp(\su_{\k}(2^{n})) = \exp(\overline{\su_{\k}(2^{n})}) \otimes U(1).$$} 
\end{notation}

\begin{theorem}{\rm The direct sum decomposition $\overline{\su_{\k}(2^{n})} = \su_{k0}(2^{n}) \oplus \su_{k1}(2^{n})$ is a Cartan decomposition of the Lie algebra $\overline{\su_{\k}(2^{n})}$.} 
\end{theorem}
{\bf Proof:} Recall that 
$$ \overline{\su_{\k}(2^{n})} = span \ \{ A\otimes I_{z},  B\otimes \Iden | A, B \in \su(2^{n-1}) \} $$
If $A, B, C \in \su(2^{n-1})$ such that $A \neq B$, then  
\begin{eqnarray*}
\left[ A\otimes I_{z}, B\otimes \Iden \right] &=& C \otimes I_{z} ;\ \\
\left[ A\otimes I_{z} , A \otimes \Iden \right] &=& 0; \\
\left[ A\otimes I_{z} , B \otimes I_{z} \right] &=& C \otimes \Iden  
\end{eqnarray*}
Therefore from the above relations and the fact that $\su(n)$ is semi-simple, it can be verified that 
\begin{eqnarray*}
\left[ \su_{\k1}(2^{n}),\su_{\k1}(2^{n}) \right] &\subset& \su_{\k0}(2^{n})\\
\left[ \su_{\k1}(2^{n}),\su_{\k0}(2^{n}) \right] &=& \su_{\k1}(2^{n})\\
\left[ \su_{\k0}(2^{n}),\su_{\k0}(2^{n}) \right] &\subset& \su_{\k0}(2^{n})
\end{eqnarray*}

\n Hence the result follows. \hfill{\bf{Q.E.D}}

\begin{corollary}\label{cor:decomp.1} {\rm Any $U \in \exp(\overline{\su_{\k}(2^n)})$ has a unique decomposition $$U = K_1 A K_2 ,$$ where $ K_1, K_2 \in SU(2^{n-1})$ and $A = \exp(Y)$, for some $Y \in \f$, a Cartan subalgebra of the pair $(\overline{\su_{\k}(2^{n})}, \su_{\k0}(2^{n}))$.} 
\end{corollary}

\n {\bf Proof:} The proof follows directly from theorem \ref{th:decomp}.

\n The above stated theorems, therefore give us a recursive decomposition procedure for an element in $SU(2^n)$. To summarize what we have accomplished till now. 

\begin{corollary} \label{cor:decomp.2} {\rm Any $U \in SU(2^n)$ has a decomposition $$U = K_1 \cdot \exp(Z_1) \cdot K_2 \cdot \exp(Y) \cdot K_3 \cdot \exp(Z_2) \cdot K_4 ,$$ where $ K_1, K_2, K_3, K_4 \in SU(2^{n-1}) \otimes U(1)$, $Y \in \h$, a Cartan subalgebra of the pair $\left(\su(2^{n}), \su_{\k}(2^{n})\right)$ and $Z_1, Z_2 \in \f$, a Cartan subalgebra of the pair $\left(\overline{\su_{\k}(2^{n})}, \su_{\k0}(2^{n})\right)$.} 
\end{corollary}

\n {\bf Proof:} The result follows directly from corollaries \ref{cor:decomp.0} and \ref{cor:decomp.1}.

\begin{remark} {\rm We know how to produce unitary transformations $K_1, K_2, K_3, K_4$ as these belong to the group $SU(2^{n-1}) \otimes U(1)$, a subgroup of $SU(2^{n-1}) \otimes SU(2)$. All that needs to be shown now is that we can produce unitary transformations of the form $\exp(Z_1)$, $\exp(Z_2)$ and $\exp(Y)$, where  $Y \in \h$, a Cartan subalgebra of the pair $\left(\su(2^{n}), \su_{\k}(2^{n})\right)$ and $Z_1, Z_2 \in \f$, a Cartan subalgebra of the pair $\left(\overline{\su_{\k}(2^{n})}, \su_{\k0}(2^{n})\right)$.This will be achieved by using the coupling between the $n^{th}$ spin and the network of coupled $n-1$ spins. We will show that the adjoint action of $SU(2^{n-1})\otimes SU(2)$ on the coupling Hamiltonian $(I_{(n-1)z}I_{nz})$ contains commuting elements whose span is the whole Cartan subalgebra $\h$. Same applies for the Cartan subalgebra $\f$ (Recall we used this fact in the two spin case, where we generated any element of the Cartan subalgebra, of the form $\alpha_1 I_{1x}I_{2x} + \alpha_2 I_{1y}I_{2y} + \alpha_3 I_{1z}I_{2z}$ by the adjoint action of $SU(2)\otimes SU(2)$ on $I_{1z}I_{2z}$). Now using the result of corollary (\ref{cor:decomp.2}) we will be able to generate any unitary transformation on $n$ qubits. } 
\end{remark}

\subsection{Generating Cartan subalgebras from Weyl orbits} 

Let us now characterize a Cartan subalgebra for the pair $\left(\su(2^{n}), \su_{\k}(2^{n})\right)$ and $\left(\overline{\su_{\k}(2^{n})}, \su_{\k0}(2^{n})\right)$, in the our product operator basis. This will be done in a recursive way. For this we use the following notation
\begin{notation} {\rm Let 
\begin{eqnarray*}
\a(2) = \pm \{iI_{1x}I_{2x}, iI_{1y}I_{2y}, iI_{1z}I_{2z} \} \\
\s(2) = \a(2)\\
\a(n) = \{\pm iI_{nx}, A \otimes I_{x} | A \in \s(n-1) \} \\
\b(n) = \{A \otimes I_{z}|A \in \s(n-1) \} \\
\s(n) = \{\s(n-1)\otimes \Iden, \a(n) \} \\
\h(n) = span\ \a(n) \\
\f(n) = span\ \b(n) 
\end{eqnarray*}}
\end{notation}

\begin{remark}{\rm To avoid confusion, we remind the reader that for $A \in \s(n-1)$, the elements $A \otimes I_{x}$ and $A \cdot I_{nx}$ represent the same object.}   
\end{remark}

\begin{example} {\rm To fix ideas, we give here the explicit expressions for $\a(n)$ and $\b(n)$ for $n=3$ and $n=4$
\begin{eqnarray*}
\a(3) &=& \pm \ i\{I_{3x}, I_{1x}I_{2x}I_{3x}, I_{1y}I_{2y}I_{3x}, I_{1z}I_{2z}I_{3x} \} \\
\a(4) &=& \pm \ i\{I_{4x}, I_{3x}I_{4x}, I_{1x}I_{2x}I_{4x}, I_{1y}I_{2y}I_{4x}, I_{1z}I_{2z}I_{4x}, I_{1x}I_{2x}I_{3x}I_{4x}, I_{1y}I_{2y}I_{3x}I_{4x}, I_{1z}I_{2z}I_{3x}I_{4x} \} \\
\b(3) &=& \pm \ i\{I_{1x}I_{2x}I_{3z}, I_{1y}I_{2y}I_{3z}, I_{1z}I_{2z}I_{3z} \} \\
\b(4) &=& \pm \ i\{I_{3x}I_{4z}, I_{1x}I_{2x}I_{4z}, I_{1y}I_{2y}I_{4z}, I_{1z}I_{2z}I_{4z}, I_{1x}I_{2x}I_{3x}I_{4z}, I_{1y}I_{2y}I_{3x}I_{4z}, I_{1z}I_{2z}I_{3x}I_{4z} \} 
\end{eqnarray*}}
\end{example}

\begin{lemma}{\rm The subspaces $\h(n)$ and $\f(n)$ are a Cartan subalgebra for the Lie algebra pair $\left( \su(2^{n}), \su_{\k}(2^{n}) \right)$ and $\left( \overline{\su_{\k}(2^{n})}, \su_{\k0}(2^{n}) \right)$ respectively.}
\end{lemma}

\n {\bf Proof:}\ We first show that $\h(n)$ is a Cartan subalgebra of  $\left( \su(2^{n}), \su_{\k}(2^{n}) \right)$. The proof is inductive. We saw that $\h(2)$ is a Cartan subalgebra of the pair $\left( \su(4), \su(2) \oplus \su(2) \right)$. We assume that the span of $\s(n-1)$ is maximally Abelian in $\su(2^{n-1})$ (easily verified for $\s(2)$). We now show $\h(n)$ is Abelian. Observe, it suffices to prove that all the elements of $\a(n)$ commute. The commutation relations $$[I_{nx}, A \cdot I_{nx}] = 0, \ \ A \in \s(n-1),$$$$[B\cdot I_{nx}, A \cdot I_{nx}] = 0, \ \ A, B \in \s(n-1),$$ and $$[B, A \cdot I_{nx}] = 0, \ \ A, B \in \s(n-1),$$ suffice to show that $\a(n)$ and $\s(n)$ are Abelian too. 

\n To show $\h(n)$ is maximally Abelian in $\su_{\m}(2^{n})$, we use induction again. Recall that $$\su_{\m}(2^{n}) = span \{ A\otimes I_{x},  A\otimes I_{y}, iI_{nx}, iI_{ny} | A \in \su(2^{n-1}) \}, $$ and $I_{nx}$ does not commute with $I_{ny},\ I_{nz}$ and $A\otimes I_{y},\ A\otimes I_{z}$, where $A \in \su(2^{n-1})$. The set $\s(n-1).I_{nx}$ is maximally Abelian in $\su(2^{n-1}) \otimes I_x$ (Because $\s(n-1)$ is maximally Abelian in $\su(2^{n-1})$ by assumption). Therefore the span of $\a(n) = \{\pm iI_{nx}, A \cdot I_{nx} | A \in \s(n-1) \}$ is maximally Abelian in $\su_m(2^{n})$ and the span of $\s(n) = \{\a(n), \s(n-1)\otimes \Iden \}$ is maximally Abelian in $\su(2^{n})$. Hence the induction argument is complete.

\n The proof $\f(n)$ is a Cartan subalgebra of the pair $\left( \overline{\su_{\k}(2^{n}}), \su_{\k0}(2^{n}) \right)$ follows exactly on same lines, and we leave it for the reader to verify it. \hfill{\bf{Q.E.D}}.

\begin{remark}{\rm We now show that we can generate any element of the Cartan Subalgebras $\h(n)$ and $\f(n)$, via the adjoint action of $SU(2^{n-1}) \otimes SU(2)$ on the coupling Hamiltonian $I_{(n-1)z}I_{nz}$. To be more precise, any element $Y \in \h$ can be written as
$$ Y = \sum_{j=1}^p \beta_j Ad_{K_j}(I_{(n-1)z}I_{nz}), \ \ \beta_j \geq 0, \ \ K_j \in SU(2^{n-1}) \otimes SU(2), $$ such that $ Ad_{K_j}(I_{(n-1)z}I_{nz})$ all commute (elements of a Weyl orbit under the adjoint action of $SU(2^{n-1}) \otimes SU(2)$ ) and therefore $$\exp (Y) = \Pi_{j =1}^p K_j \cdot \exp(\beta_j I_{(n-1)z}I_{nz})\cdot K_j^{\dagger}.$$ }
\end{remark}

\begin{lemma}{\rm Let $K = SU(2^{n-1}) \otimes SU(2)$, then the adjoint orbit $Ad_K(I_{(n-1)z}I_{nz})$ contains the sets $\s(n-1)\otimes I_x$ and $\s(n-1)\otimes I_z = \b(n)$.} 
\end{lemma}

\n {\bf Proof:} The proof is again inductive. Note by definition
$$\s(n-1)\otimes I_z = \{ \s(n-2) \otimes \Iden \otimes I_z, \ \s(n-2) \otimes I_x \otimes I_z , \ \pm iI_{(n-1)x} I_{nz} \}.$$ We assume that the statement of the lemma is true for $n-1$, i.e., if $H = SU(2^{n-2}) \otimes SU(2)$, then $Ad_H(I_{(n-2)z}I_{(n-1)z})$ contains the set $\s(n-2) \otimes I_{x}$. Therefore the adjoint orbit $Ad_K(I_{(n-2)z} I_{(n-1)z} I_{nz})$ contains the set $\s(n-2) \otimes I_{x}\otimes I_{z}$ (Apply $H$ to $I_{(n-2)z}I_{(n-1)z}$ and don't do anything to last spin (qubit)!). Now observe for $$U_1 = \exp(- i \pi I_{(n-2)z} I_{(n-1)y})$$ and $$U_2 = \exp(i \frac{\pi}{2} I_{(n-1)y})$$ both belonging to $K$, we have $$U_2 U_1 (I_{(n-1)z} I_{nz}) U_1^{\dagger} U_2^{\dagger} = I_{(n-2)z} I_{(n-1)z} I_{nz}.$$ Since the term $\pm I_{(n-1)x} I_{nz}$ are present in adjoint orbit $Ad_K(I_{(n-1)z} I_{nz})$, (By selective $\frac{\pi}{2}$ rotation of $n-1$ qubit) we deduce that the adjoint orbit $Ad_K(I_{(n-1)z} I_{nz})$ contains the set $$\a(n-1) \otimes I_{z} = \{\s(n-2) \otimes I_x \otimes I_z , \ \pm I_{(n-1)x} I_{nz} \}.$$ Finally observe terms of the form $\s(n-2) \otimes \Iden \otimes I_z$ can be produced by adjoint action of $K_1 = \exp(-i \pi I_{(n-1)x}I_{ny})$ on the set $\s(n-2) \otimes I_{x} \otimes I_{x}$, which produces Hamiltonians of the form  $\s(n-2) \otimes \Iden \otimes I_{z}$. Thus we can generate the whole set $\s(n-1) \otimes I_{z}$ . That we can also generate $\s(n-1) \otimes I_{x}$, is also obvious (by selective $\frac{\pi}{2}$ rotation of the $n^{th}$ spin). \hfill{\bf{Q.E.D}}.

\begin{remark}{\rm Note that the Hamiltonian $\pm I_{nx}$ can be produced by selective excitation of the $n^{th}$ qubit. Thus we can produce any Hamiltonian from the set $\{ \s(n-1) \otimes I_x , \pm I_{nx} \}= \a(n)$. We have therefore shown that we can produce any Hamiltonian $Y$ and $Z$ belonging to the set $\a(n)$ and $\b(n)$ and as there positive span is the Cartan subalgebra $\h(n)$ and $\b(n)$ we can generate any Hamiltonian belonging to $\h(n)$ and $\b(n)$ by the adjoint action of the group $SU(2^{n-2}) \otimes SU(2)$ on the coupling Hamiltonian and we are done. We complete the inductive proof with the following lemma.}
\end{remark}  

\begin{lemma}\label{lm:3spin.cartan}{\rm Let $K = SU(4) \otimes SU(2)$, then the adjoint orbit $Ad_K(I_{2z}I_{3z})$ contains the sets $\a(2)\otimes I_x = \s(2)\otimes I_x$ and $\a(2)\otimes I_z = \b(3)$.} 
\end{lemma}

\n {\bf Proof:} Recall
\begin{eqnarray*}
\a(3) &=& \pm \ i\{I_{3x}, I_{1x}I_{2x}I_{3x}, I_{1y}I_{2y}I_{3x}, I_{1z}I_{2z}I_{3x} \} \\
\b(3) &=& \pm \ i\{I_{1x}I_{2x}I_{3z}, I_{1y}I_{2y}I_{3z}, I_{1z}I_{2z}I_{3z} \} \\
\end{eqnarray*} 
The result then follows from the following relations
\begin{eqnarray*}
\exp(- i\pi I_{1x}I_{2y}) \exp(-i\pi \alpha I_{2z}I_{3z}) \exp(i \pi I_{1x}I_{2y}) &=& \exp(-i \pi \alpha I_{1x}I_{2x}I_{3z}) \\
\exp(-i \frac{\pi}{2} I_{1z}) \exp(-i \frac{\pi}{2} I_{2z}) \exp(-i\pi \alpha I_{1x}I_{2x}I_{3x}) \exp(i \frac{\pi}{2} I_{2z}) \exp(i \frac{\pi}{2} I_{1z}) &=& \exp(-i \pi \alpha I_{1y}I_{2y}I_{3z}) \\  
\exp(-i \frac{\pi}{2} I_{1y}) \exp(-i \frac{\pi}{2} I_{2y}) \exp(-i\alpha \pi I_{1x}I_{2x}I_{3x}) \exp(i \frac{\pi}{2} I_{2y}) \exp(i \frac{\pi}{2} I_{1y}) &=& \exp(-i \pi \alpha I_{1z}I_{2z}I_{3z}) \\
\end{eqnarray*} which shows how to generate $\b(3)$. Now terms in $\a(2)\otimes I_{x}$ can be generated by selectively rotating the third spin.  \hfill{\bf{Q.E.D}}.

To review the ideas developed and to conclude the paper, let us explicitly work out the synthesis of any unitary propogator in a linearly scalar coupled three spin system.

\begin{example}{\rm Consider three heteronuclear spin $\frac{1}{2}$ nuclei, coupled through scalar coupling . One has the ability to selectively excite each of the nuclei. Thus the system evolution in a suitable rotating frame is approximated by \begin{equation} \label{eq:het.two} \dot{U} = -i (\ H_d + \sum_{i=1}^{6}u_i H_i \ )U , \end{equation} where \begin{eqnarray*}H_d &=& 2\pi J ( I_{1z} I_{2z} + I_{2z}I_{3z} )\\
H_1 &=& 2\pi I_{1x} \\
H_2 &=& 2\pi I_{1y} \\
H_3 &=& 2\pi I_{2x} \\
H_4 &=& 2\pi I_{2y}\\
H_5 &=& 2\pi I_{3x} \\
H_6 &=& 2\pi I_{3y}
\end{eqnarray*}
Observe that the subgroup generated by $\{H_i \}_{i=1}^{6}$ is $SU(2)\otimes SU(2) \otimes SU(2)$. Thus any unitary propogator $U$ can be decomposed as

$$ U = K_1 \exp(-i ( \beta_1 I_{1x}I_{2x}I_{3x} + \beta_2 I_{1y}I_{2y}I_{3x} + \beta_3 I_{1z}I_{2z} I_{3x} + \beta_4 I_{3x})) K_2 $$ where
$$ K_1 = P_1 \exp(-i ( \alpha_1 I_{1x}I_{2x}I_{3z} + \alpha_2 I_{1y}I_{2y}I_{3z} + \alpha_3 I_{1z}I_{2z}I_{3z} )) P_2 $$ and
$$ K_2 = P_3 \exp(-i ( \gamma_1 I_{1x}I_{2x}I_{3z} + \gamma_2 I_{1y}I_{2y}I_{3z} + \gamma_3 I_{1z}I_{2z} I_{3z} )) P_4, $$ 

where $P_1, P_2, P_3, P_4$ belong to $SU(4)\otimes SU(2)$, with $SU(4)$ denoting arbitrary transformation on spin $1$ and $2$ and $SU(2)$ representing selective local transformations on the spin $I_3$. 

Any element of $SU(4)$ can be produced by decoupling the spin $I_3$ (rapidly flipping the spin \cite{Ernst}) and then using the coupling $I_{1z}I_{2z}$ as in example \ref{ex:2spin}. The Hamiltonians belonging to the Cartan subalgebra $\h(3)$ and $\f(3)$ can then be produced by using the coupling $I_{2z}I_{3z}$ as illustrated in lemma \ref{lm:3spin.cartan}.} 
\end{example}

\begin{remark}{\rm We have therefore demonstrated constructive controllability in a network of $\n$ coupled spins. The crucial thing to note is that we have dealt with the worst case scenario. The case we have treated here is of a network of linearly coupled spins. There was no direct coupling between the $n^{th}$ spin and say $k^{th}$ spin for $k < n$ (In terms of quantum computing, if $k < n-1$, we are not allowed to let $k^{th}$ and $n^{th}$ qubit interact with a two qubit gate directly). However interactions can be mediated through the other spins and this is what our constructive procedure is doing. It is also not difficult to see now that as long as we have a network of spins which is connected we have constructive controllability.}\end{remark}

\begin{center}\section{Conclusions and Future Work}\end{center}
Our main goal in this paper has been to put the design of quantum computers and pulse sequences in NMR on a geometrical footing. We have produced here a parameterization of the group $SU(2^n)$ in terms of unitary transformations produced by one and two qubit gates using successive Cartan decompositions of $SU(2^n)$.  The first issue we would like to draw attention to is time optimality. We would like to emphasize that our constructive procedure of producing an arbitrary unitary transformation is only time optimal for the two spin case \cite{time:khan}. Recall that in the course of synthesizing a propogator almost all of the time is spent during the interaction of qubits and this corresponds to the two qubit gates or evolution of couplings during NMR. The one qubit operations can be produced relatively fast by external selective hard pulses. Thus from a practical viewpoint, it is of utmost importance that not only do we have a constructive procedure for synthesizing an arbitrary unitary transformation, but one which is time optimal to minimize the effects of decoherence which are always present. In \cite{time:khan}, we developed the theory for time optimal control of spin systems and computed the time optimal control laws for two spin system (as in example \ref{ex:2spin}) for any kind of coupling between the spins. The two spin case is special and elegant because $\frac{SU(4)}{SU(2)\otimes SU(2)}$ happens to be a Riemannian symmetric space. The results of time optimality for two spin systems do not extend in a natural way to higher spin systems as the coset space $\frac{SU(2^n)}{[SU(2)]^{\otimes n}}$ does not have a symmetric space structure. Finding time optimal control laws for spin networks with more that two spins will require and inspire further developments in geometric control theory.

\n Another question of imminent interest is -- how to transform between different parameterizations of $SU(2^n)$. Suppose $U = SU(2)$ , then we can express any element of $SU(2)$ in the following two ways 
\begin{eqnarray}
U = \exp(-i \alpha_1 I_x + \alpha_2 I_y + \alpha_3 I_z) ,\ \ \alpha_1, \alpha_2, \alpha_3 \in \Re \\
U = \exp(-i \beta_1 I_x) \exp(-i \beta_2 I_y) \exp(-i \beta_3 I_x), \ \ \beta_1, \beta_2, \beta_3 \in \Re. 
\end{eqnarray}
It is well known, how to transform between these and other ways of expressing an element of $SU(2)$. It will be very interesting to find the transformation that takes our description of $SU(2^n)$, which is in the same spirits as the last of the above descriptions, to the other standard parameterizations of $SU(2^n)$.

\bibliography{\Bi/stabilization}
\bibliography{\Bi/NMR}

\begin{thebibliography}{99}

\bibitem{nomizu}  S. Kobayashi, and K. Nomizu.
{\it Foundations of Differential Geometry Vol.1 and 2} (Interscience Publishers) (1969).

\bibitem{Helg}  S. Helgason
{\it Differential Geometry, Lie Groups, and Symmetric Spaces } (Academic Press) (1978).

\bibitem{wolf}  J A. Wolf,
{\it Spaces of Constant Curvature} (Publish or Perish, Inc., 1984).

\bibitem{jurd}  V. Jurdjevic and H. J. Sussmann
{\it Journal of Differential Equations} {\bf 12}, 313 (1972).

\bibitem{Ernst}  R. R. Ernst,  G. Bodenhausen, A. Wokaun,
{\it Principles of Nuclear Magnetic Resonance in One and Two
Dimensions} (Oxford University Press, Oxford, 1987).

\bibitem{Science}

S. J. Glaser, T. Schulte-Herbr\"uggen, M. Sieveking, O. Schedletzky, N. C.
Nielsen, O.
W. S\o rensen and C. Griesinger, {\it Science} {\bf 280}, 421 (1998).

\bibitem{optics}

W. S. Warren, H. Rabitz, M. Dahleh, {\it Science} {\bf 259}, 1581 (1993).


\bibitem{QC}

N. A. Gershenfeld and I. L. Chuang, {\it Science} {\bf 275}, 350 (1997);

D. G. Cory, A. Fahmy,  T. Havel,  {\it Proc. Natl. Acad. Sci. USA}   {\bf
94}, 1634  (1997).

\bibitem{design}
T. Untidt, T. Schulte-Herbr\"uggen, B. Luy, S. J. Glaser,
C. Griesinger, O. W. S\o rensen and  N. C. Nielsen,
{\it Molecular Physics} {\bf 95}, 787 (1998);

T. Untidt, S. J. Glaser, C. Griesinger and  N. C. Nielsen,
{\it Molecular Physics} {\bf 96}, 1739 (1999).

\bibitem{transfer}
J. Cavanagh, A. G. Palmer III, P. E. Wright, M. Rance,
{\it J. Magn. Reson.} {\bf 91}, 429
(1991);

A. G. Palmer III, J. Cavanagh, P. E. Wright, M. Rance, {\it ibid.} {\bf
93}, 151 (1991);

L. E. Kay; {\it J. Am. Chem. Soc.} {\bf 115}, 2055 (1993);

M. Sattler, P. Schmidt, J. Schleucher, O. Schedletzky, S. J. Glaser
and C. Griesinger,
{\it J. Magn. Reson.} {\bf B 108}, 235 (1995);

J. Schleucher, M. Schwendinger, M. Sattler, P. Schmidt, O. Schedletzky,
S. J. Glaser, O. W. S\o rensen and C. Griesinger,
{\it J. Biomol. NMR} {\bf 4}, 30 (1994).

\bibitem{Isotropic}

D. P. Weitekamp,  J. R. Garbow,  A. Pines,
 {\it J. Chem. Phys.} {\bf 77}, 2870 (1982), ibid. {\bf 80}, 1372 (1984);

P. Caravatti,  L. Braunschweiler,  R. R. Ernst,
{\it Chem. Phys. Lett.} {\bf 100}, 305  (1983);

S. J. Glaser and J. J. Quant, in
{\it Advances in Magnetic
and Optical Resonance},
W. S. Warren, Ed. ( Academic Press, New York, 1996),
vol.  19, pp. 59-252.

\bibitem{sorenson}
O.W. S{\o}rensen, G.W. Eich, M.H. Levitt, G.Bodenhausen and R.R. Ernst,
{\it Progr. NMR Spectrosc.} {\bf 16}, 163 (1983).

\bibitem{hermes}
G.W. Haynes and H. Hermes,
{\it SIAM J. Control and Optimization} {\bf 8}, 450
(1970).

\bibitem{khaneja}
R.W. Brockett and Navin Khaneja,
``Stochastic Control of Quantum Ensembles''
{\it System Theory: Modeling, Analysis and Control} (Kluwer Academic Publishers, Inc., 1999).

\bibitem{khaneja.thesis}
Navin Khaneja,
{\it Geometric Control in Classical and Quantum Systems} (Ph.d. Thesis, Harvard University, 2000).

\bibitem{judson}
R. S. Judson, K. K. Lehmann, H. Rabitz, W. S. Warren,
{\it J. Mol. Struct.} {\bf 223}, 425
(1990).

\bibitem{herbruggen}
T. Schulte-Herbr\"uggen
{\it Aspects and Prospects of High-Resolution NMR} (Ph.d. Thesis, ETH Zurich, 1998).

\bibitem{shor}P.W. Shor, ``Algorithms for quantum computation: discrete log and factoring'', Proceedings of the 35th Annual Symposium on the Foundations of Computer Science (IEEE Computer Society Press, Los Alamitos, CA, 1994), p.124.

\bibitem{grover}L.K. Grover, ``Quantum mechanics helps in searching for a needle in a haystack'' {\it Phys. Rev. Lett.} {\bf 79}, 325-328

\bibitem{deutsch} D.Deutsch, ``Quantum Computational networks'',{\it Proc. Roy. Soc. Lond. A}{\bf 425}, 73 (1989).

\bibitem{lloyd} S.Lloyd, ``Almost any quantum logic gate is universal'', {\it Phys. Rev. Lett.}, {\bf 75}, 346-349

\bibitem{barenco.et.al}A. Barenco, C.H. Bennett, R. Cleve, D.P. Divincenzo, N. Margolus, P. Shor, T. Sleator, J. Smolin and H. Weinfurter, ``Elementary gates for quantum computation'', {\it Physical Review A}, {\bf 28}, 1996 

\bibitem{Divincenzo} D.P. DiVincenzo, ``Two-bit gates are universal for quantum computation'', Phys. Rev. A {\bf 50}, 1015 (1995).

\bibitem{rama} V. Ramakrishna, M.V. Salapaka, M. Dahleh, H. Rabitz, and A. Peirce, Phys. Rev. A {\bf 51}, 960 (1995)

\bibitem{tarn} G.M. Huang, T.J. Tarn, and J.W. Clark, J. Math. Phys. {\bf 24} , 2608 (1983)

\bibitem{gilmore} Robert Gilmore, {\it Lie Groups, Lie algebras and Some of Their
Applications}, (Wiley-Interscience, 1974).

\bibitem{time:khan} Navin Khaneja, Roger Brockett, and Steffen. J. Glaser, ``Time optimal control of spin systems'', quant-ph/0006114, {\it submitted to Physical Review A}.

\end{thebibliography}

\end{document}